\documentclass[reqno,12pt,draft]{article}
\usepackage{amssymb,euscript}

\newcommand{\e}{\begin{equation}}
\newcommand{\ee}{\end{equation}}
\newcommand{\ea}{\begin{eqnarray}}
\newcommand{\eea}{\end{eqnarray}}
\newcommand{\nn}{\nonumber\\}
\newcommand{\p}{\partial}

\newcommand{\Tr}{{\rm Tr}}
\newcommand{\A}{{\cal A}}
\newcommand{\C}{\mathbb C}

\begin{document}

\begin{flushright}
\end{flushright}
\begin{flushright}
\end{flushright}
\begin{center}

{\LARGE {\sc Gauge Symmetry and\\ Supersymmetry of Multiple \\
M2-Branes \\}}

\bigskip
{\sc Jonathan Bagger\footnote{bagger@jhu.edu}}\\
{Department of Physics and Astronomy\\
Johns Hopkins University\\
3400 North Charles Street\\
Baltimore, MD 21218, USA }
\\
\bigskip
and
\bigskip
\\
{\sc Neil Lambert\footnote{neil.lambert@kcl.ac.uk}} \\
{Department of Mathematics\\
King's College London\\
The Strand\\
London WC2R 2LS, UK\\}

\end{center}

\bigskip
\begin{center}
{\bf {\sc Abstract}}
\end{center}

In previous work we proposed a field theory model for multiple
M2-branes based on an algebra with a totally antisymmetric triple
product.  In this paper we gauge a symmetry that arises from the
algebra's triple product.  We then construct a supersymmetric theory
that is consistent with all the symmetries expected of a multiple
M2-brane theory: 16 supersymmetries, conformal invariance, and
an $SO(8)$ R-symmetry that acts on the eight transverse scalars.  
The gauge field is not dynamical.  The result is a new
type of maximally\break supersymmetric gauge theory in three
dimensions.

\newpage

\section{\sl Introduction}

The branes of M-theory are important but still very much mysterious
objects.  While the dynamics of a single M-brane is well understood,
very little is known about the interactions of multiple M-branes.
For a current review of M-branes and their interactions, see
\cite{Berman:2007bv}.

In a recent paper \cite{Bagger:2006sk}, we proposed a model
of multiple M2-branes based on an algebra that admits a
totally antisymmetric triple product.  (The triple product can be
constructed, for example, from the associator in a nonassociative
algebra.)  Examination of the supersymmetry algebra
suggested that the theory has a local gauge symmetry
that arises from the triple product.

In ref.\,\cite{Bagger:2006sk} the nature of these gauge
transformations was not clear, so the model presented
contained just the scalar and fermi fields.  Moreover, it
was invariant under just four supersymmetries.  In this
paper we will study the gauge symmetry in more detail.
We will show how to gauge the local symmetry and obtain
a conformal and gauge-invariant action with all 16
supersymmetries.  The theory an $SO(8)$ R-symmetry
that acts on the eight transverse scalars, a nonpropagating
gauge field, and no free parameters, modulo a rescaling
of the structure constants.  The gauge field
ensures that the supersymmetry
algebra closes (up to a gauge transformation) on shell.

Apart from our motivation to obtain a worldvolume theory for
multiple M2-branes, it is generally worthwhile to to pursue
extensions to Yang-Mills gauge theory and to explore the possible
relevance of nonassociative structures to theoretical physics and
geometry.  In fact, beyond the model proposed in
\cite{Bagger:2006sk}, there are other physical examples
\cite{Englert:1987fm}--\cite{Papageorgakis:2006ed} in which fuzzy
three-spheres arise. Such objects presumably require an algebraic
structure that is based on an antisymmetric triple product, so the
results discussed here may be relevant.

The rest of this paper is organized as follows. In section two we
present the details of the algebraic structure that we require and
show how it leads to a natural symmetry. In section three we gauge
the symmetry by introducing a vector gauge field. In section four we
construct a gauge-invariant supersymmetric theory with 16
supercharges acting on the scalars, vector and fermions.  The
superalgebra closes on a set of equations of motion that are
invariant under supersymmetry.  We show that the equations of motion
arise from a supersymmetric action consistent with all the known
continuous symmetries of the M2-brane.  Section five contains some
closing comments.

We also include two appendices. The first provides a concrete
example of a three-algebra; the second lists some Clifford algebra
identities that are relevant to the computations in section four.

\bigskip

While this paper was is in preparation, we received
ref.\,\cite{Gustavsson:2007vu}, in which the algebraic structures
underlying multiple M2-branes are discussed. Furthermore, in a
revised version (v4), the gauged supersymmetry algebra was found to
close using the fermion and vector equations of motion. The fields 
are elements of a Lie algebra constructed out of
the semidirect product of two other algebras, one of which has a
triple product. The superalgebra presented in
\cite{Gustavsson:2007vu} looks similar to ours.  It would be
interesting to see if the two algebraic structures are, in fact, the
same.

\section{\sl Some Algebraic Details}

The model presented in \cite{Bagger:2006sk} was based on
a nonassociative algebra.  In algebra one commonly
introduces the associator
\begin{equation}
<A,B,C>\  =\, (A\cdot B)\cdot C - A\cdot (B\cdot C),
\end{equation}
which vanishes in an associative algebra.  In what follows
we need the antisymmeterized associator
\begin{eqnarray} [A,B,C] &=& <A,B,C> + <B,C,A>+<C,A,B>\\
\nonumber &&-<A,C,B>-<B,A,C>-<C,B,A>,
\end{eqnarray}
which is what one finds by expanding out the Jacobi identity
$[[A,B],C]+[[B,C],A]+[[C,A],B]$.  In a nonassociative algebra,
the antisymmeterized associator leads to a natural triple product structure.

To define an action we require a trace-form on the algebra $\A$.
This is a bilinear map $\Tr: \A\times\A\to \C$ that is symmetric
and invariant:
\begin{equation}
\Tr(A,B)=\Tr(B,A)\qquad \Tr(A\cdot
B,C) = \Tr(A,B\cdot C).
\end{equation}
We also assume `Hermitian' conjugation $\#$ and positivity, which
implies $\Tr(A^\#,A)\ge 0$ for any $A\in \A$
(with equality if and only if $A=0$). The invariance property
implies that
\begin{eqnarray}
\Tr(<A,B,C>,D)&=& \Tr((A\cdot B)\cdot C,D)-\Tr(A\cdot (B\cdot
C),D)\nn &=& \Tr(A\cdot B,C\cdot D)-\Tr(A,(B\cdot C)\cdot D)\\ &=&
-\Tr(A,<B,C,D>)\nonumber .
\end{eqnarray}
It also follows that
\begin{equation}
\Tr([A,B,C],D) =-\Tr(A,[B,C,D]). \label{invprop}
\end{equation}
More generally we only require that the algebra admit a totally
antisymmetric trilinear product $[\cdot,\cdot,\cdot]$ that
satisfies (\ref{invprop}). In particular, the antisymmetric product
need not arise from a non-associative product on the algebra.  We
call such an algebra a three-algebra. Note that a three-algebra need
not contain a bilinear product and hence is not necessarily an
algebra in the usual sense.

In \cite{Bagger:2006sk} we found that closure of the 16
component supersymmetry algebra leads to the variation
\begin{equation}
\delta X^I \propto i\bar\epsilon_2\Gamma_{JK}\epsilon_1[X^J,X^K,X^I],
\label{gaugedsym}
\end{equation}
which can be viewed as a local version of the global symmetry
transformation
\begin{equation}
\delta X = [\alpha,\beta,X], \label{trans}
\end{equation}
where $\alpha,\beta\in \A$.  For (\ref{trans}) to
be a symmetry, it must act as a derivation,
\begin{equation}
\delta([X,Y,Z])= [\delta X,Y,Z]+[X,\delta Y,Z]+[X,Y,\delta Z].
\end{equation}
This leads to the `fundamental' identity (which has also appeared in
\cite{Okubo}--\cite{Pioline:2002ba})
\begin{equation}
[\alpha,\beta,[X,Y,Z]]=[[\alpha,\beta,X],Y,Z]+[X,[\alpha,\beta,Y],Z]
+[X,Y,[\alpha,\beta,Z]].
\label{fundamental}
\end{equation}
We proceed assuming that this identity holds. It will play
a role analogous to the Jacobi identity in ordinary Lie algebra,
where it arises from demanding that the transformation
$\delta X = [\alpha,X]$ act as a derivation.

It is convenient to introduce a basis $T^a$ for the algebra $\A$. On
physical grounds we assume that all the generators are Hermitian, in
the sense that $(T^a)^\#=T^a$.  We then expand the field $X =
X_aT^a$, $a = 1,..., N$, where $N$ is the dimension of $\cal A$ (and
not the number of M2-branes). We introduce the `structure' constants
\begin{equation}
[T^a,T^b,T^c] = f^{abc}{}_{d}T^d,
\end{equation}
from which is it is clear that $f^{abc}{}_{d}=f^{[abc]}{}_{d}$. The
trace-form provides a metric
\begin{equation}
h^{ab} = \Tr(T^a,T^b)
\end{equation}
that we can use to raise indices: $f^{abcd}=h^{de}f^{abc}{}_{e}$.
Again on physical grounds we assume that $h^{ab}$ is
positive definite.  The condition (\ref{invprop}) on the
trace-form implies that
\begin{equation}
f^{abcd} = -f^{dbca},
\end{equation}
and this further implies that
$f^{abcd} =f^{[abcd]}$, in analogy with the familiar result in
Lie algebras. In a basis form the fundamental identity
(\ref{fundamental}) becomes
\begin{equation}
f^{efg}{}_{d}f^{abc}{}_{g}=f^{efa}{}_{g}f^{bcg}{}_{d}
+f^{efb}{}_{g}f^{cag}{}_{d}+f^{efc}{}_{g}f^{abg}{}_{d}.
\end{equation}
We can augment this algebra by including an element $T^0$
that associates with everything, or more precisely, that satisfies
$f^{0ab}{}_{d}=0$.  If we assume that $h^{0b}=0$ if $b\ne 0$,
we find $f^{abc}{}_{0}=0$. Thus this mode decouples and
it can be interpreted as the centre-of-mass coordinate.

The symmetry transformation (\ref{trans}) can be written as
\begin{equation}
\delta X_d = f^{abc}{}_{d}\alpha_a\beta_b X_c.
\end{equation}
However the notation allows for the more general transformation
\begin{equation}
\delta X_d = f^{abc}{}_{d}\Lambda_{ab}X_c,
\end{equation}
which we assume from now on.  In particular, the transformation
(\ref{gaugedsym}) corresponds to the choice
\begin{equation}
\Lambda_{ab} \propto i\bar\epsilon_1\Gamma_{JK}\epsilon_2X^J_aX^K_b.
\end{equation}
Note that the generator $\Lambda_{ab}$ cannot in general be written
as $\alpha_{[a}\beta_{b]}$ for a single pair of vectors ($\alpha_a,
\beta_b$).  However, $\Lambda_{ab}$ can always be written as a
sum over $N$ such pairs.

To see that the action is invariant under global symmetries of this
form, we observe that for any $Y$,
\begin{eqnarray}\label{lengths}
\frac{1}{2}\delta \Tr(Y,Y)&=& \Tr(\delta Y,Y)\nn &=&
h^{de}\delta Y_d Y_e\\
&=& h^{de}\Lambda_{ab} f^{abc}{}_{d}Y_cY_e\nn &=&
f^{abce}\Lambda_{ab}Y_cY_e\nn &=&0,\nonumber
\end{eqnarray}
by the antisymmetry of $f^{abce}$.  In addition, the fundamental
identity ensures that
\begin{equation}
(\delta [X^I,X^J,X^K])_a = f^{cdb}{}_{a}
\Lambda_{cd}[X^I,X^J,X^K]_b.
\end{equation}
Thus the Lagrangian
\begin{equation}
{\cal L}=-\frac{1}{2}\Tr(\p_\mu X^I,\p^\mu X^I) -
3\kappa^2\Tr([X^I,X^J,X^K],[X^I,X^J,X^K]),
\end{equation}
is invariant under the symmetry $\delta X^I_a = f^{cdb}{}_{a}
\Lambda_{cd}X^I_b$.

\section{\sl Gauging the Symmetry}

We now wish to promote the global symmetry discussed above
to a local one.  To that end we introduce a covariant derivative
$D_\mu X$ such that $\delta (D_\mu X) = D_\mu (\delta X)
+ (\delta D_\mu) X$. If we let
\begin{equation}
\delta X_a = \Lambda_{cd}f^{cdb}{}_{a}X_b \equiv \tilde
\Lambda^b{}_{a}X_b,
\end{equation}
then the natural choice is to take
\begin{equation}
(D_\mu X)_a = \p_\mu X_a - \tilde A_{\mu}{}^b{}_a X_b,
\end{equation}
where $\tilde A_{\mu}{}^{b}{}_a\equiv f^{cdb}{}_{a}A_{\mu cd}$ is a
gauge field with two algebraic
indices. We can therefore think of $\tilde
A_{\mu}{}^b{}_a$ as living in the space of linear maps from $\cal A$
to itself, in analogy with the adjoint representation of a Lie
algebra. The field $X$ is then, in some sense, in the
fundamental representation.  The gauge field acts as an element of
$gl(N)$, where $N$ is the dimension of $\cal A$. Furthermore, as a
consequence of the antisymmetry of $f^{abcd}$, the symmetry algebra
is contained in $so(N)$.

A little calculation shows that the covariant derivative is obtained by
taking
\begin{eqnarray}
\nonumber \delta \tilde A_{\mu}{}^b{}_a &=&\p_\mu \tilde
\Lambda^b{}_{a} -\tilde \Lambda^b{}_{c}\tilde A_{\mu}{}^c{}_a +
\tilde A_{\mu}{}^b{}_c \tilde \Lambda^c{}_a\\
&\equiv & D_\mu \tilde \Lambda^b{}_a.
\end{eqnarray}
Indeed, this is the usual form of a gauge transformation.  The field strength is defined as
\begin{equation}
([D_\mu,D_\nu]X)_a = \tilde F_{\mu\nu}{}^b{}_a X_b,
\end{equation}
which leads to
\begin{eqnarray}
\tilde F_{\mu\nu}{}^b{}_a  &=&\p_\nu \tilde A_{\mu}{}^b{}_a  -
\p_\mu \tilde A_{\nu}{}^b{}_a-\tilde A_{\mu}{}^b{}_c\tilde A_{\nu}{}^c{}_a
+ \tilde A_{\nu}{}^b{}_c \tilde A_{\mu}{}^c{}_a .
\end{eqnarray}
The resulting Bianchi identity is $D_{[\mu} \tilde F_{\nu\lambda]}{}^b{}_a=0$.
One also finds that
\begin{equation}
\delta \tilde F_{\mu\nu}{}^b{}_a
=-\tilde \Lambda^b{}_{c} \tilde F_{\mu\nu}{}^c{}_a + \tilde F_{\mu\nu}{}^b{}_c \tilde \Lambda^c{}_a.
\end{equation}
These expressions are identical to what one finds in an ordinary
gauge theory based on a Lie algebra, where the gauge field is in the
adjoint representation.  Here the gauge field takes values in the
space of linear maps of $\cal A$ into itself.  The triple product
allows one to construct linear maps on $\cal A$ from two elements of
$\cal A$.

In particular consider the set $\cal G$ of all $N\times N$ matrices
$\tilde \Lambda^b{}_{a} = \Lambda_{cd}f^{cdb}{}_a$, where
$\Lambda_{cd}$ is arbitrary. The fundamental identity ensures that
this set is closed under the ordinary matrix commutator. Thus $\cal
G$ defines a matrix Lie algebra that is a subalgebra of $so(N)$.
The fundamental identity implies that $f^{abcd}$ is
an invariant 4-form of $\cal G$. Thus every three-algebra generates
a Lie algebra with an invariant 4-form. However, it is unclear
whether or not the existence of an invariant 4-form in a Lie algebra
is sufficient to ensure that its fundamental representation is a
three-algebra that satisfies the fundamental identity.

\section{\sl Supersymmetrizing the Gauged Theory}

We now show how to supersymmeterize the gauged
multi-M2-brane model in a manner consistent will all the
continuous symmetries expected of a multiple M2-brane
theory, namely
16 supersymmetries, conformal invariance, and an $SO(8)$
R-symmetry that acts on the eight transverse scalars.  We
first recall the structure of the full superalgebra with 16-component
spinors. In \cite{Bagger:2006sk} we argued that the general
form is
\begin{eqnarray}\label{susy}
\delta X^I &=& i\bar\epsilon\Gamma^I\Psi\\
\nonumber \delta \Psi &=& \p_\mu X^I\Gamma^\mu \Gamma^I\epsilon
+\kappa[X^I,X^J,X^K]\Gamma^{IJK}\epsilon,
\end{eqnarray}
where  $\kappa$ is a constant.  We then showed that this
algebra does not close.  However, closure on the scalars $X^I$
leads to the local symmetry
\begin{equation}
\delta X^I \propto
i \bar\epsilon_2\Gamma_{JK}\epsilon_1[X^J,X^K,X^I]
\end{equation}
that we gauged above.

Let us apply the ideas of the previous section to gauge this
symmetry.  We start by introducing the gauge field $\tilde
A_{\mu}{}^b{}_a$ with its associated covariant derivative.
The supersymmetry transformations then take the form
\begin{eqnarray}\label{susygauged}
\nonumber \delta X^I_a &=& i\bar\epsilon\Gamma^I\Psi_a\\
\delta \Psi_a &=& D_\mu X^I_a\Gamma^\mu \Gamma^I\epsilon +\kappa
X^I_bX^J_cX^K_d f^{bcd}{}_{a}\Gamma^{IJK}\epsilon \\
\nonumber \delta\tilde A_{\mu}{}^b{}_a &=& i\bar\epsilon
\Gamma_\mu\Gamma_IX^I_c\Psi_d f^{cdb}{}_{a}.
\end{eqnarray}
(A similar, possibly identical, form for the gauge field variation
was used in \cite{Gustavsson:2007vu}.)

This algebra can be made to close on shell.  We first consider the
scalars.  We find that the transformations close into a translation
and a gauge transformation;
\begin{equation}\label{Sclose}
[\delta_1,\delta_2]X^I_a =
 v^\mu D_\mu X^I_a +
\tilde\Lambda^b{}_a X^I_b
\end{equation}
where
\begin{equation}\label{vLambdaDef}
v^\mu = -2i\bar\epsilon_2\Gamma^\mu\epsilon_1,\qquad
\tilde\Lambda^b{}_a =6i\kappa \bar\epsilon_2\Gamma_{JK}\epsilon_1
X^J_c X^K_d f^{cdb}{}_{a} .
\end{equation}

We then consider the fermions.  Evaluating $[\delta_1,\delta_2]\Psi_a$,
we find two separate terms involving
$\bar\epsilon_2\Gamma_{\mu}\Gamma_{IJKL}\epsilon_1$
that must cancel for closure.  This implies
\begin{equation}\label{betagamma}
\kappa = -1/6,
\end{equation}
so there is no free parameter. Proceeding further we compute
\begin{eqnarray}
\nonumber  [\delta_1,\delta_2]\Psi_a &=&  v^\mu D_\mu \Psi_a +
\tilde\Lambda^b{}_a \Psi_b \\
  && +\ i(\bar\epsilon_2\Gamma_\nu\epsilon_1)\Gamma^\nu\left(\Gamma^\mu D_\mu\Psi_a
+\frac{1}{2}\Gamma_{IJ}X^I_cX^J_d\Psi_b f^{cdb}{}_a\right)\\
  \nonumber &&-\   \frac{i}{4}(\bar\epsilon_2\Gamma_{KL}\epsilon_1)\Gamma^{KL}\left(\Gamma^\mu D_\mu\Psi_a
+\frac{1}{2}\Gamma_{IJ}X^I_cX^J_d\Psi_b f^{cdb}{}_a\right).
\end{eqnarray}
Closure requires that the second and third lines vanish.  This
determines the fermionic equation of motion;
\begin{equation}\label{Feq}
\Gamma^\mu D_\mu\Psi_a
+\frac{1}{2}\Gamma_{IJ}X^I_cX^J_d\Psi_bf^{cdb}{}_{a}=0.
\end{equation}
Thus on shell we see that
\begin{equation}
[\delta_1,\delta_2]\Psi_a  = v^\mu D_\mu\Psi_a + \tilde\Lambda^b{}_a \Psi_b,
\end{equation}
as required.

We finally turn to $[\delta_1,\delta_2]\tilde A_\mu{}^b{}_a$. Here
we again find a term involving $\bar\epsilon_2\Gamma_{\mu}\Gamma_{IJKL}\epsilon_1$:
\begin{equation}
-\frac{i}{3}(\bar\epsilon_2\Gamma_{\mu}\Gamma_{IJKL}\epsilon_1)
X^I_c
X^J_eX^K_fX^L_gf^{efg}{}_{d}f^{cdb}{}_{a}.
\end{equation}
Happily this term vanishes as a consequence of the fundamental
identity. Continuing, we find
\begin{eqnarray}
  \nonumber [\delta_1,\delta_2]\tilde A_\mu{}^b{}_a &=&  2i(\bar\epsilon_2\Gamma^\nu\epsilon_1)
  \epsilon_{\mu\nu\lambda} (X^I_cD^\lambda X^I_d + \frac{i}{2}\bar\Psi_c\Gamma^\lambda\Psi_d)f^{cdb}{}_{a} \\
   && -\ 2i(\epsilon_2\Gamma_{IJ}\epsilon_1)X^I_cD_\mu X^J_df^{cdb}{}_{a}.
\end{eqnarray}
To close the algebra we fix the $\tilde A_{\mu}{}^b{}_a$ equation of
motion;
\begin{equation}\label{AEq}
 \tilde F_{\mu\nu}{}^b{}_a
   +\epsilon_{\mu\nu\lambda}(X^J_cD^\lambda X^J_d
+\frac{i}{2}\bar\Psi_c\Gamma^\lambda\Psi_d )f^{cdb}{}_{a}  = 0,
\end{equation}
so that on shell,
\begin{equation}
\nonumber [\delta_1,\delta_2]\tilde A_\mu{}^b{}_a = v^\nu \tilde
F_{\mu\nu}{}^b{}_a   + D_\mu\tilde\Lambda^b{}_a.
\end{equation}
Note that $\tilde A_{\mu\ d}^{\ c}$ contains no local
degrees of freedom, as required. We see that the 16
supersymmetries close on shell.

To find the bosonic equations of motion, we take the supervariation
of the fermion equation of motion.  This gives
\begin{eqnarray}
  \nonumber 0&=& \Gamma^I\left( D^2X^I_a-\frac{i}{2}\bar\Psi_c\Gamma^{IJ}X^J_d\Psi_b f^{cdb}{}_{a}
  +\frac{1}{2}f^{bcd}{}_{a}f^{efg}{}_{d}X^J_bX^K_cX^I_eX^J_fX^K_g \right)\epsilon\\
  && +\ \Gamma^I\Gamma_\lambda X^I_b\left(\frac{1}{2}\varepsilon^{\mu\nu\lambda}\tilde F_{\mu\nu}{}^b{}_a
  - X^J_cD^\lambda X^J_df^{cdb}{}_{a} -\frac{i}{2}\bar\Psi_c\Gamma^\lambda\Psi_d f^{cdb}{}_{a}\right)
  \epsilon.
\end{eqnarray}
The second term vanishes as a consequence of the vector equation of
motion (\ref{AEq}).  The first term determines the scalar equations of
motion,
\begin{eqnarray}\label{Xeq}
D^2X^I_a-\frac{i}{2}\bar\Psi_c\Gamma^{IJ}X^J_d\Psi_b f^{cdb}{}_a
  -\frac{\partial V}{\partial X^{Ia}}   &=& 0.
\end{eqnarray}
The potential is
\begin{eqnarray}\label{potential}
\nonumber V  &=& \frac{1}{12}f^{abcd}f^{efg}{}_d
X^I_aX^J_bX^K_cX^I_eX^J_fX^K_g \\
 &=& \frac{1}{2\cdot 3!}{\rm Tr}([X^I,X^J,X^K],[X^I,X^J,X^K]).
\end{eqnarray}

Let us summarize our results. The supersymmetry transformations are
\begin{eqnarray}\label{susygauged2}
\nonumber \delta X^I_a &=& i\bar\epsilon\Gamma^I\Psi_a\\
\delta \Psi_a &=& D_\mu X^I_a\Gamma^\mu \Gamma^I\epsilon -\frac{1}{6}
X^I_bX^J_cX^K_d f^{bcd}{}_{a}\Gamma^{IJK}\epsilon \\
\nonumber \delta\tilde A_{\mu}{}^b{}_a &=& i\bar\epsilon
\Gamma_\mu\Gamma_IX^I_c\Psi_d f^{cdb}{}_{a}.
\end{eqnarray}
These supersymmetries close into translations and
gauge transformations,
\begin{eqnarray}
  \nonumber [\delta_1,\delta_2] X^I_a &=&  v^\mu\partial_\mu X^I_a
  +(
\tilde\Lambda^b{}_a-v^\nu\tilde A_\nu{}^b{}_a X^I_b )\\
 {}[\delta_1,\delta_2] \Psi_a  &=& v^\mu \partial_\mu\Psi_a + (\tilde\Lambda^b{}_a
 -v^\nu\tilde A_\nu{}^b{}_a \Psi_b )\\
  \nonumber [{\delta_1},{\delta_2}] \tilde A_\mu{}^b{}_a &=& v^\nu\partial_\nu  \tilde
 A_{\mu}{}^b{}_a +\tilde D_\mu(\tilde
\Lambda^b{}_a -v^\nu\tilde A_\nu{}^b{}_a ),
\end{eqnarray}
after using the equations of motion
\begin{eqnarray}\label{EOMS}
\nonumber\Gamma^\mu D_\mu\Psi_a
+\frac{1}{2}\Gamma_{IJ}X^I_cX^J_d\Psi_bf^{cdb}{}_{a}&=&0\\
 D^2X^I_a-\frac{i}{2}\bar\Psi_c\Gamma^I_{\ J}X^J_d\Psi_b f^{cdb}{}_a
   -\frac{\partial V}{\partial X^{Ia}}    &=& 0 \\
\nonumber \tilde F_{\mu\nu}{}^b{}_a
  +\varepsilon_{\mu\nu\lambda}(X^J_cD^\lambda X^J_d
+\frac{i}{2}\bar\Psi_c\Gamma^\lambda\Psi_d )f^{cdb}{}_{a}  &=& 0.
\end{eqnarray}
We have explicitly demonstrated that the supersymmetry variation
of the fermion equation of motion vanishes, and that the algebra closes
on shell.  It follows that all the equations of motion are invariant
under supersymmetry.
Furthermore one can check using the fundamental identity that the
Bianchi identity $\epsilon^{\mu\nu\lambda} D_{\mu}\tilde
F_{\nu\lambda}{}^b{}_a =0$ is satisfied.

We close this section by presenting an action for this system.
The equations of motion can be obtained from the Lagrangian
\begin{eqnarray}\label{action}
\nonumber {\cal L} &=& -\frac{1}{2}(D_\mu X^{aI})(D^\mu X^{I}_{a})
+\frac{i}{2}\bar\Psi^a\Gamma^\mu D_\mu \Psi_a
+\frac{i}{4}\bar\Psi_b\Gamma_{IJ}X^I_cX^J_d\Psi_a f^{abcd}\\
&& - V+\frac{1}{2}\varepsilon^{\mu\nu\lambda}(f^{abcd}A_{\mu
ab}\partial_\nu A_{\lambda cd} +\frac{2}{3}f^{cda}{}_gf^{efgb}
A_{\mu ab}A_{\nu cd}A_{\lambda ef}).
\end{eqnarray}
It is not hard to check that the action is gauge invariant and
supersymmetric under the transformations (\ref{susygauged2}).
Note that (\ref{action}) contains no free parameters, up to a
rescaling of the structure constants.  In fact, given the presence
of the Chern-Simons term, it is natural to expect the $f^{abcd}$
to be quantized \cite{BLinprep}.

It is important to note that the structure constants $f^{abc}{}_{d}$
enter into the
Chern-Simons term in a non-standard way.  Viewed as a 3-form
in an arbitrary dimension, this `twisted' Chern-Simons term
\begin{equation}
\Omega =(f^{abcd}A_{\mu ab}\partial_\nu A_{\lambda cd}
+\frac{2}{3}f^{cda}{}_gf^{efgb} A_{\mu ab}A_{\nu cd}A_{\lambda
ef})\, dx^\mu\wedge dx^\nu \wedge dx^\lambda
\end{equation}
satisfies
\begin{equation}
d\Omega = F_{ab}\wedge \tilde F^{ab},
\end{equation}
where $\tilde F_{\mu\nu}{}^b{}_a = F_{\mu\nu cd}f^{cdb}{}_a$.
Also note that $\Omega$ is written in
terms of $A_{\mu ab}$ and not the physical field $\tilde
A_{\mu}{}^b{}_a = A_{\mu cd}f^{cdb}{}_a$ that appears in the
supersymmetry transformations and equations of motion. However,
one can check that $\Omega$ is invariant under shifts of
$A_{\mu ab}$ that leave $\tilde A_{\mu}{}^b{}_a$ invariant. Thus
it is locally well defined as a function of $\tilde A_{\mu}{}^b{}_a$.

This theory provides an example of the type of model that was
searched for in \cite{Schwarz:2004yj}.  It is invariant under 16
supersymmetries and an $SO(8)$ R-symmetry.  It is also conformally
invariant at the classical level. These are all the continuous
symmetries that are expected of multiple M2-branes.  Note that the
Chern-Simons term naively breaks the parity that is expected to be a
symmetry of the M2-brane worldvolume. However, we can make the
Lagrangian parity invariant if we assign an odd parity to $f^{abcd}$.  In
particular, if we invert $x^2 \to -x^2$, we must then require
that  $X^I_a$ and $\tilde A^{\ a}_{\mu\ b}$ be parity even for $\mu =
0,1$; $ \tilde A^{\ a}_{2\ b}$ and $f^{abcd}$ be parity odd; and
$\Psi_a\to \Gamma_2\Psi_a$.  Note that this assignment implies that
$A_{\mu ab}$ is parity odd for $\mu=0,1$, while $A_{2 ab}$ is
parity even.

\section{\sl Conclusions}

In this paper we described the gauge symmetry that arises in the
model of multiple M2-branes presented in \cite{Bagger:2006sk}.  We
included a nonpropagating gauge field and obtained a theory that is
invariant under all 16 supersymmetries with no free parameters,
up to a rescaling. Thus
the model presented in \cite{Bagger:2006sk} can indeed be viewed
as the truncation of a maximally supersymmetric theory to the scalar
and fermion modes.

The Lagrangian given here is consistent with all the known
symmetries of M2-branes. The M2-brane worldvolume theory
is expected to arise as the strong coupling, conformal fixed
point of a three-dimensional, maximally supersymmetric
Yang-Mills gauge theory.
Furthermore, in the large $N$ limit, it is conjectured to be
dual to an $AdS_4\times S^7$ solution of M-theory.  Thus
the Lagrangian given here is a candidate for the strong
coupling fixed point of a three-dimensional super
Yang-Mills theory and a field theory dual of M-theory on
$AdS_4\times S^7$.

As mentioned in the introduction, similar results on the
closure of the algebra have recently been reported in
\cite{Gustavsson:2007vu}.  This paper adopts a different, but
possibly equivalent, form for the algebra.  We hope that these
studies will warrant a deeper and fruitful investigation into the
algebraic structure of multiple M2-branes as a step towards
identifying the microscopic degrees of freedom of M-theory.

\section*{\sl Acknowledgements}

We would like to thank S.\,Cherkis and N.\,Cook for discussions. We
also thank A.\,Gustavsson for alerting us to the revised version
(v4) of \cite{Gustavsson:2007vu}.  JB is supported in part by the US
National Science Foundation, grant NSF-PHY-0401513. NL is supported
in part by the PPARC grant PP/C507145/1 and the EU grant
MRTN-CT-2004-512194 and would like to thank the Isaac Newton
Institute where this work was completed.

\section*{\sl Appendix A: An Example}

In this appendix we provide an example of a three-algebra that
satisfies the fundamental identity. The simplest nontrivial case
corresponds to four generators, $a,b,...=1,2,3,4$.  If we
normalize the generators such that ${\rm Tr}(T^a,T^b)\propto
\delta^{ab}$, it then follows that
\begin{equation}
f^{abcd}\propto \varepsilon^{abcd}.
\end{equation}
One can explicitly check that the fundamental identity is satisfied.
In this case the space $\cal G$ generated by all matrices $\tilde
\Lambda^c{}_d=\Lambda_{ab}f^{abc}{}_d$ is the space of all $4\times
4$ anti-symmetric matrices and hence ${\cal G} = so(4)$ with the
invariant 4-form $\varepsilon^{abcd}$.

It is also possible to realize this three-algebra as arising from a
non-associative algebra. In \cite{Bagger:2006sk} we considered the
three-algebra of Hermitian matrices that anti-commute with a fixed
Hermitian matrix $G$, with $G^2=1$. We defined
\begin{equation}
A\cdot B = QABQ,
\end{equation}
where $Q=\frac{1}{\sqrt2}(1+iG)$. We also took ${\rm Tr}(A,B)
= {\rm trace}(Q^{-1}AQ^{-1}B)$ where trace denotes the
standard matrix trace. The associator turned out to be
\begin{equation} <A,B,C> =
2GABC,
\end{equation}
and hence
\begin{equation}
[A,B,C] = 2G(ABC\pm {\rm cyclic}).
\end{equation}
We also found that $G$ could play the role of translations as
$[A,B,G]=0$ for all $A,B$ that anticommute with $G$.

Let us consider the case in which $\cal A$ has
four generators, which we take to be the (Euclidean)
four-dimensional $\gamma$-matrices with $G=\gamma_5$
and $Q = (1+i\gamma_5)/\sqrt{2}$. The product is then
$\gamma^a\cdot\gamma^b = Q\gamma^a\gamma^bQ =
\gamma_5\gamma^a\gamma^b$, and one finds that
\begin{equation}
<\gamma^a,\gamma^b,\gamma^c> =  2 \gamma_5\gamma^a\gamma^b\gamma^c.
\end{equation}
Thus
\begin{equation}
[\gamma^a,\gamma^b,\gamma^c] =  2\cdot 3! \gamma_5\gamma^{abc} =
2\cdot 3!\varepsilon^{abc}{}_{d}\gamma^d,
\end{equation}
and hence $f^{abcd}=12\varepsilon^{abcd}$.

\section*{\sl Appendix B: Fierz and Other Identities}

In this appendix we present the Fierz identity that we use
repeatedly above.  All spinorial quantities are those of the
eleven-dimensional Clifford algebra; we take them to be real.
Let $\epsilon_1,\epsilon_2$ and $\chi$ be arbitrary spinors.
The combination $(\bar\epsilon_2\chi)\epsilon_1 -
(\bar\epsilon_1\chi)\epsilon_2$ can then be written as
\begin{eqnarray}\label{fierz}
 &&(\bar\epsilon_2\chi)\epsilon_1 -
(\bar\epsilon_1\chi)\epsilon_2 =\\
\nonumber&&-\frac{1}{16}\left((\bar\epsilon_2\Gamma_m\epsilon_1)\Gamma^m\chi
-\frac{1}{2!}(\bar\epsilon_2\Gamma_{mn}\epsilon_1)\Gamma^{mn}\chi
+\frac{1}{5!}(\bar\epsilon_2\Gamma_{mnpqr}\epsilon_1)\Gamma^{mnpqr}\chi,
\right),
\end{eqnarray}
where $m,n,...=0,...,10$, $\mu,\nu,... = 0,1,2$ and $I,J,...=
3,4,...,10$. If $\epsilon_1$ and $\epsilon_2$ have the same
chirality with respect to  $\Gamma_{012}$, then the only terms
that contribute must have an even number of $I$ indices.
Moreover, the expression is only nonvanishing when $\chi$
has the same $\Gamma_{012}$ chirality as $\epsilon_1$ and
$\epsilon_2$.  When this is the case, (\ref{fierz}) reduces to
\begin{eqnarray}\label{fierzplus}
 &&(\bar\epsilon_2\chi)\epsilon_1 -
(\bar\epsilon_1\chi)\epsilon_2 =\\
\nonumber&&-\frac{1}{16}\left(2(\bar\epsilon_2\Gamma_\mu\epsilon_1)\Gamma^\mu\chi
-(\bar\epsilon_2\Gamma_{IJ}\epsilon_1)\Gamma^{IJ}\chi
+\frac{1}{4!}(\bar\epsilon_2\Gamma_{\mu}\Gamma_{IJKL}\epsilon_1)\Gamma^\mu\Gamma^{IJKL}\chi
\right).
\end{eqnarray}

We also found the following identities useful:
\begin{eqnarray}
  \nonumber \Gamma_M\Gamma^{IJ}\Gamma^M&=& 4\Gamma^{IJ}\\
\nonumber  \Gamma_M\Gamma^{IJKL}\Gamma^M &=&0\\
\nonumber \Gamma^{IJP}\Gamma^{KLMN}\Gamma_P&=&-\Gamma^I\Gamma^{KLMN}\Gamma^J +\Gamma^J\Gamma^{KLMN}\Gamma^I \\
 \nonumber  \Gamma^I\Gamma^{KL}\Gamma^J - \Gamma^J\Gamma^{KL}\Gamma^I &=& 2\Gamma^{KL}\Gamma^{IJ}  - 2\Gamma^{KJ}\delta^{IL}
+ 2\Gamma^{KI}\delta^{JL}- 2\Gamma^{LI}\delta^{JK}\\
\nonumber&& + 2\Gamma^{LJ}\delta^{IK}- 4\delta^{KJ}\delta^{IL}+4\delta^{KI}\delta^{JL}   \\
  \nonumber \Gamma^{IJM}\Gamma^{KL}\Gamma_M &=&2\Gamma^{KL}\Gamma^{IJ}  - 6\Gamma^{KJ}\delta^{IL}
+ 6\Gamma^{KI}\delta^{JL}- 6\Gamma^{LI}\delta^{JK}\\
&& + 6\Gamma^{LJ}\delta^{IK}+ 4\delta^{KJ}\delta^{IL}-4\delta^{KI}\delta^{JL}.
\end{eqnarray}

\end{document}